\documentclass[journal]{IEEEtran}

\usepackage{verbatim}
\usepackage[noadjust]{cite}

\ifCLASSINFOpdf
  \usepackage[pdftex]{graphicx}
\else
\fi
\usepackage{amssymb,amsmath,amsthm,bm}
\usepackage{amsfonts}

\def\vt{{\bm{t}}}

\def\vw{{\bm{w}}}
\def\vx{{\bm{x}}}
\def\vy{{\bm{y}}}

\usepackage[colorlinks = true,
            linkcolor = blue,
            urlcolor  = blue,
            citecolor = blue,
            anchorcolor = blue]{hyperref}

\begin{document}

\title{Deep Interactive Denoiser (DID) for X-Ray Computed Tomography}

\author{Ti~Bai,
        Biling~Wang,
        Dan~Nguyen,
        Bao~Wang,
        Bin~Dong,
        Wenxiang~Cong,
        Mannudeep~K.~Kalra,
        and~Steve~Jiang
        \vspace{-2.8em}

\thanks{Ti Bai, Dan Nguyen, Biling Wang and Steve Jiang (corresponding author) are with the Medical Artificial Intelligence and Automation (MAIA) Laboratory, Department of Radiation Oncology, University of Texas Southwestern Medical Center, Dallas, Texas 75239, USA (email: \mbox{Ti.Bai@UTSouthwestern.edu}, \mbox{Dan.Nguyen@UTSouthwestern.edu}, \mbox{Biling.Wang@UTSouthwestern.edu}, \mbox{Steve.Jiang@UTSouthwestern.edu}).

Bao Wang is with the Department of Mathematics and Scientific Computing and Imaging Institute, University of Utah, Salt Lake City, UT, USA (email: \mbox{bwang@math.utah.edu})

Bin Dong is with the Beijing International Center for Mathematical Research, Peking University, Beijing, China (email: \mbox{dongbin@math.pku.edu.cn})

Wenxiang Cong is with the Department of Biomedical Engineering, Rensselaer Polytechnic Institute, Troy, NY, 12180, USA. (email: \mbox{wxcong@gmail.com})

Mannudeep K. Kalra is with the Department of Radiology, Massachusetts General Hospital and Harvard Medical School, Boston, MA, USA (email: \mbox{mkalra@mgh.harvard.edu})
}
}
\maketitle

\begin{abstract}
Low-dose computed tomography (LDCT) is desirable for both diagnostic imaging and image-guided interventions. Denoisers are widely used to improve the quality of LDCT. Deep learning (DL)-based denoisers have shown state-of-the-art performance and are becoming mainstream methods. However, there are two challenges to using DL-based denoisers: 1) a trained model typically does not generate different image candidates with different noise-resolution tradeoffs, which are sometimes needed for different clinical tasks; and 2) the model's generalizability might be an issue when the noise level in the testing images differs from that in the training dataset. To address these two challenges, in this work, we introduce a lightweight optimization process that can run on top of any existing DL-based denoiser during the testing phase to generate multiple image candidates with different noise-resolution tradeoffs suitable for different clinical tasks in real time. Consequently, our method allows users to interact with the denoiser to efficiently review various image candidates and quickly pick the desired one; thus, we termed this method deep interactive denoiser (DID). Experimental results demonstrated that DID can deliver multiple image candidates with different noise-resolution tradeoffs and shows great generalizability across various network architectures, as well as training and testing datasets with various noise levels.
\end{abstract}


\section{Introduction}
\label{sec: introduction}
\IEEEPARstart{M}{edical} imaging techniques play vital roles in healthcare for diagnosing disease and guiding clinical procedures. However, the clinical value of these techniques may be substantially decreased if the medical images they produce do not faithfully reflect clinical realities because of image quality degradations. Specifically, quantum noise is one of the dominant factors that degrades X-ray computed tomography (CT) because this application demands low doses of radiation. A strong quantum noise might overwhelm potential low-contrast lesions, which would make it challenging to diagnose disease or guide procedures. Therefore, a denoiser is usually required to enhance the image quality by suppressing the noise effectively while preserving the image resolution as much as possible. 

Denoising is a longstanding problem in the X-ray CT imaging field, and, accordingly, it has been researched abundantly \cite{RN1,RN2,RN3,RN4,RN5,RN6,RN7,RN8,RN9,RN10,RN11,RN12,RN13,RN14,RN15,RN16,RN17,RN18,RN19,RN20,RN21,RN22}. In recent years, deep learning (DL) techniques have seen unprecedented success in many fields \cite{RN23,RN24,RN25,RN26,RN27,RN28,RN29,RN30,RN31,RN32,RN33,RN34,RN35} and have also been widely employed for medical image denoising \cite{RN1,RN4,RN12,RN13,RN20,RN22}. For instance, Chen \textit{et al.} built a residual encoder-decoder convolutional neural network (CNN) to enhance the quality of low-dose CT (LDCT) images  \cite{RN10}. Shan \textit{et al.} demonstrated a 3D CNN for volumetric CT image denoising, which can be pretrained from a 2D version \cite{RN22}. Yang \textit{et al.} used a generative adversarial network (GAN) to denoise CT images \cite{RN13}. Li \textit{et al.} proposed a multi-stage network to gradually improve enhanced CT images \cite{RN12}. Since deep learning methods can automatically learn powerful features directly from an existing database, which usually exhibit stronger expressiveness compared to models manually designed by human experts, these models' denoising performance is superior to that of conventional methods. Accordingly, deep learning methods have become mainstream. 

Despite the great progress, several problems still must be solved before DL-based denoisers can be widely deployed in real clinical practice. First, model generalizability is a major problem. Theoretically speaking, to maximize denoising performance, DL-based denoisers should be used to denoise images that have the same noise levels as the training dataset. In other words, a denoiser can only reach its maximum performance when the training and testing domains match exactly. For example, to have the optimal denoising performance for CT images acquired at a certain low-dose level, the desired network should be trained by feeding the LDCT images with the same low-dose level into the network to predict the full-dose level CT images. One solution would be to train multiple models on different training datasets to aim at different noise levels. However, one can hardly consider all the noise levels in model training because, in clinical practice, there are too many factors that contribute to the noise level, such as different anatomical sites, equipment vendors, and imaging protocols. 

Second, it is well accepted that medical image quality is highly task-specific, and sometimes observer-specific, because different tasks require different noise-resolution tradeoffs, and different observers with different clinical backgrounds have their own tradeoff preferences even for the same clinical tasks. Therefore, a good practical denoiser should be able to produce multiple images with different noise-resolution tradeoffs so that users can pick the ones they desire. This can be achieved relatively easily with a conventional regularized iterative algorithm \cite{RN2,RN6,RN7,RN9,RN15,RN16,RN17,RN18,RN19,RN21,RN36,RN50}, where  regularization parameters can be adjusted to generate a noise-resolution tradeoff (NRT) curve consisting of denoised images with various noise-resolution tradeoffs to meet various clinical demands. However, because of their black-box nature, most DL-based denoisers do not have this clinically significant feature.

To the best of our knowledge, little effort has been made to solve these two clinically important problems regarding DL-based CT denoisers. The most relevant work came from Shan \textit{et al.}, who proposed a modularized adaptive processing neural network (MAP-NN) \cite{RN20}. In that work, the authors proposed a unique training framework that can modulate the denoising direction by designing a module-clone-based network architecture. Once the MAP-NN has been well trained, each single module can suppress the noise at a small step. Thereby, the trained denoiser can produce multiple images to meet the user’s preferences.

Since many groups have developed DL-based CT denoisers that can deliver state-of-the-art results, it would be of great value to solve the two problems above by adding a light plug-in to these denoisers. In this study, our purpose is to develop such a light plug-in to empower existing DL-based CT denoisers to 1) allow users to interactively and continuously adjust the noise-resolution tradeoff and 2) adapt to new testing images that have different noise levels from the training images. The proposed method, which we have termed deep interactive denoiser (DID), is part of our efforts to develop human-centered artificial intelligence (AI) in medicine. We believe that, in many challenging clinical scenarios, AI should assist human experts in making clinical decisions or performing clinical procedures better and faster, not replace human experts by fully automating everything. In the proposed LDCT denoising applications, DID will assist users to interactively tune images to achieve the noise-resolution tradeoff desired for the specific task and the particular user. It should be noted that DID is directly used in the model testing phase and does not require a training stage, so it can be readily combined with any existing DL-based denoiser.

\section{Methods and Materials}
\subsection{Methods}
\label{sec:methods}

Let us first mathematically formulate the X-ray CT denoising problem. Basically, given a noisy image $\vx \in \mathrm{R}^N$, the denoising task is to restore the underlying clean counterpart $\vy \in \mathrm{R}^N$, where the images are vectorized representations and $N$ denotes the number of pixels. If the noise $\bm{\epsilon} \in \mathrm{R}^N$ is assumed to be additive, we have:
    \vspace{-0.3cm}
\begin{equation}
    \vx = \vy + \bm{\epsilon}.
\end{equation}

The above problem can be effectively solved in a learning-and-prediction fashion. More specifically, in the context of DL-based denoising methods, a CNN $\phi_\vw$ parameterized by $\vw$ can be learned from a training dataset $\{(\vx_m,\vy_m)|m \in (0,1,\cdots,M-1)\}$ with a size of $M$, where $m$ indexes the noisy-clean image pairs, by minimizing the following cost function:
\begin{equation}
    \label{eq:supervised_training}
    \vw_0 = \arg \min_{\vw} \sum_{m=0}^{M-1} ||\phi_\vw(\vx_m) - \vy_m||_2^2.
\end{equation}

Once the network training (learning) process is finished, a denoised image $\bar{\vy}$ can be predicted by feeding a noisy image $\vx$ into the network:
    \vspace{-0.3cm}
\begin{equation}
    \label{eq:default_image}
    \bar \vy = \phi_{\vw_0}(\vx).
\end{equation}

From~(\ref{eq:default_image}), regarding the conventional model deployment, the learned CNN $\phi$ is a one-to-one mapping if the learned network parameters $\vw_0$ are fixed (which is also a common and default setting in practice) during the model testing phase.

One of the purposes of this study is to equip any existing DL-based denoiser with the features to dynamically produce denoised images with the desired noise-resolution tradeoff from among all the achievable tradeoffs. To achieve this purpose, we propose a method to generate multiple denoised images with different noise-resolution tradeoffs based on a denoised image  $\bar{\vy}$  produced by a pre-trained DL-based denoiser $\phi_{\vw_0}$ for a given LDCT image $\vx$.

\begin{figure}[tb!]
    \centering
    \includegraphics[width=0.3\textwidth]{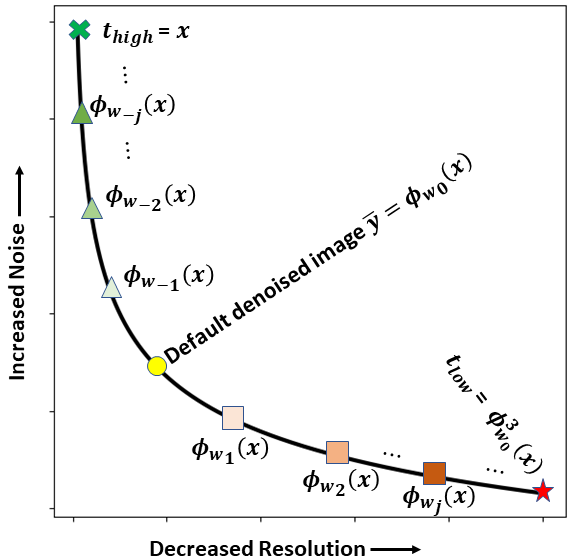}
    \caption{Illustration of the main idea of the proposed DID algorithm. We assume that all denoised images with different noise-resolution tradeoffs fall on a noise-resolution tradeoff curve. The default denoised image $\bar{\vy}=\phi_{\vw_0}\left(\vx\right)$ (yellow circle) is generated by a pre-trained DL-based denoiser. Bound point $\vt_{\mathrm{high}}$ corresponds to an image with high resolution and high noise (green cross), and bound point point $\vt_{\mathrm{low}}$ corresponds to an image with low resolution and low noise (red star). Two bound points are determined that bracket most, if not all, images with clinically meaningful noise-resolution tradeoffs. The images between $\vt_{\mathrm{high}}$ and $\bar{\vy}$ (triangles) and the images between $\bar{\vy}$ and $\vt_{\mathrm{low}}$ (squares) were all generated with the proposed DID algorithm.}
    \label{fig:pareto}
        \vspace{-0.8cm}

\end{figure}
Figure~\ref{fig:pareto} illustrates the main idea of the proposed DID algorithm. We assume that all denoised images with different noise-resolution tradeoffs can construct a noise-resolution tradeoff (NRT) curve, and that most, if not all, images denoised from an LDCT image $\vx$ with clinically meaningful noise-resolution tradeoffs can be bracketed within two bound points, $\vt_{\mathrm{high}}$  and $\vt_{\mathrm{low}}$, where $\vt_{\mathrm{high}}$  corresponds to a denoised image with high resolution and high noise and $\vt_{\mathrm{low}}$ corresponds to a denoised image with low resolution and low noise. The default denoised image $\bar{\vy}=\phi_{\vw_0}\left(\vx\right)$ generated by a pre-trained DL-based denoiser $\phi_{\vw_0}$ falls between $\vt_{\mathrm{high}}$  and $\vt_{\mathrm{low}}$. The goal of the proposed DID algorithm is to populate this NRT curve between $\vt_{\mathrm{high}}$  and $\vt_{\mathrm{low}}$ by using the three known points: $\bar{\vy}$, $\vt_{\mathrm{high}}$, and $\vt_{\mathrm{low}}$.

In this work, we chose the original noisy LDCT image $\vx$ as $\vt_{\mathrm{high}}$, which represents the highest resolution and highest noise, as it is very unlikely that someone would want to add more noise to the original noisy image. To generate $\vt_{\mathrm{low}}$, we purposely overly denoise x by recursively applying the pre-trained DL-based denoiser $\phi_{\vw_0}$ for $K$ times: $\vt_{\mathrm{low}}=\phi_{\vw_0}^K\left(\vx\right)=\phi_{\vw_0}\left(\cdots\left(\phi_{\vw_0}\left(\vx\right)\right)\right)$, where $K$ is determined empirically such that it is the smallest number that can still ensure that $\vt_{\mathrm{low}}$  has the lowest resolution and lowest noise of any image with clinically meaningful noise-resolution tradeoffs. In this work, we found $K=3$. 

To populate the NRT curve between $\vt_{\mathrm{high}}$ and $\vt_{\mathrm{low}}$, we will fine-tune the pre-trained denoiser $\phi_{\vw_0}$ by solving the following problems:
\begin{equation}
    \vw = \arg \min_{\vw} ||\phi_\vw(\vx) - \vt_{\mathrm{low}}||_2^2,
    \label{eq:towards_low_noise}
\end{equation}

\begin{equation}
    \vw = \arg \min_{\vw} ||\phi_\vw(\vx) - \vt_{\mathrm{high}}||_2^2.
    \label{eq:towards_high_resolution}
\end{equation}

Problems~(\ref{eq:towards_low_noise}) and~(\ref{eq:towards_high_resolution}) can be efficiently iteratively solved with the gradient descent algorithm. During the optimization process, a series of network parameters $\vw_0,\vw_{\pm 1},\cdots,\vw_{\pm J}$ from different iterations can be generated, where $\vw_0$ is the starting point with respect to the pre-trained denoising network $\phi_{\vw_0}$. Here, we use $\vw_1,\cdots,\vw_J$ to denote images generated by solving problem~(\ref{eq:towards_low_noise}), which have lower resolution and lower noise than the default denoised image $\bar{\vy}=\phi_{\vw_0}\left(\vx\right)$. We use $\vw_{-1},\cdots,\vw_{-J}$ to denote images generated by solving problem~(\ref{eq:towards_high_resolution}), which have higher resolution and higher noise than the default denoised image $\bar{\vy}=\phi_{\vw_0}\left(\vx\right)$. 

\begin{figure*}
    \centering
    \includegraphics[width=0.8\textwidth]{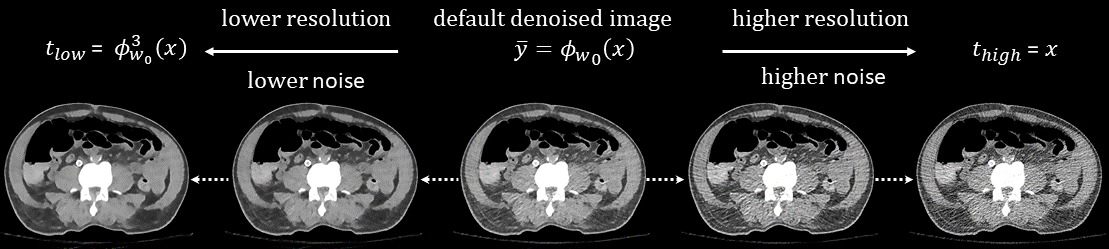}
        \vspace{-0.3cm}
    \caption{Gradually transiting noise-resolution tradeoffs among images generated by the proposed DID algorithm. The middle image is the default denoised image $\bar{\vy}=\phi_{\vw_0}\left(\vx\right)$. From the middle to the right, images with higher resolution are generated by solving problem~(\ref{eq:towards_high_resolution}). From the middle to the left, images with lower noise are generated by solving problem~(\ref{eq:towards_low_noise}). Display window: [-160, 240] HU.}
    \label{fig:flowchart}
    \vspace{-0.5cm}
\end{figure*}

It is well accepted that deep neural networks can easily fit a wide variety of functions, given their large model capacity. Since minimizing cost function~(\ref{eq:towards_low_noise}) in the testing phase is equivalent to training a network with a training dataset consisting of only one sample $(\vx,\vt_{\mathrm{low}})$, this “one sample” training dataset can very easily be overfitted. As a result, it is expected that $\phi_{\vw_J}\left(\vx\right)=\vt_{\mathrm{low}}$ when the optimization process is converged. During this optimization process, we observed that the network’s output with updated model parameters ${\vw_1,\cdots,\vw_J}$ will gradually transit from the initial image $\phi_{\vw_0}\left(\vx\right)$ to the final target image $\vt_{\mathrm{low}}=\phi_{\vw_J}\left(\vx\right)$. Basically, $\vt_{\mathrm{low}}$ guides the network parameters to be updated in a specified direction towards lower noise and lower resolution, such that the associated network output ${\phi_{\vw_1}\left(\vx\right),\cdots,\phi_{\vw_J}\left(\vx\right)}$ will populate the NRT curve between $\bar{\vy}$ and $\vt_{\mathrm{low}}$ and exhibit noise-resolution tradeoffs with gradually lower resolution and lower noise. The reasoning for populating the NRT curve between $\vt_{\mathrm{high}}$ and  $\bar{\vy}$ by minimizing cost function~(\ref{eq:towards_high_resolution}) is similar. Figure~\ref{fig:flowchart} provides a simple illustration about this process. 

In this work, we would also provide a heuristic argument to further show that the intermediate results $\phi_{\vw_{\pm 1}}(\vx), \phi_{\vw_{\pm 2}}(\vx)\cdots\phi_{\vw_{\pm J}}(\vx)$ exhibit monotonic noise and resolution changes.

Let $L(\vw):=\|\phi_{\vw}(\vx)- \vt\|_2^2$, for any given pair $(\vx, \vt)$. Consider gradient flow $$\vw_{j+1}=\vw_j-\eta\nabla L(\vw).$$By definition, we have that $L(\vw_j)$ monotonically decreases when the learning rate $\eta$ is small enough. When $\phi_{\vw}$ is over-parameterized, then by the theory of neural tangent kernel (NTK) \cite{jacot2018neural}, $L(\vw_{j})\to0$ exponentially as $j\to\infty$.

It was recently discovered that, in the NTK regime \cite{RN547}, the training dynamics minimizing $L(\vw)$ with gradient flow can be written as
$$\phi^{j+1}=\phi^j-\eta\Theta_{\vx}(\phi^j-t),$$ where we define $\phi^j=\phi_{\vw_j}(\vx)$ for clarity. This dynamic is identical to the procedure known as twicing \cite{RN548}, except that the choice of kernel $\Theta_{\vx}$ is different. 

Suppose that $\vx$ is a noisy image and $\vt$ is a low resolution and low noise image (e.g., $\vt_{\mathrm{low}} = \phi_{\vw_0}^3(\vx)$). Then, the algorithm starts with $\phi^0=\phi_{\vw_0}(\vx)$, and $\phi^0-\vt=\phi_{\vw_0}(\vx)-\phi_{\vw_0}^3(\vx)=\delta_1$ represents the noise (as well as some sharp features) that is still present in $\phi_{\vw_0}(\vx)$, which is gradually removed during the iteration. Similarly, suppose that $\vt$ is a high resolution and high noise image (e.g., $\vt_{\mathrm{high}} = \vx$ itself). Then, the algorithm starts with $\phi^0=\phi_{\vw_0}(\vx)$, and $\phi^0-\vt=\phi_{\vw_0}(\vx)-\vx=-\delta_2$ represents the sharp features (as well as noise) that is removed from $\phi_{\vw_0}(\vx)$, which is gradually injected back during the iteration.

As such, our DID algorithm can now use the bound images to guide the modification of the model parameters from the pre-trained parameters $\vw_0$ towards two different directions with respect to noise-resolution tradeoffs. We expect that 1) the images generated towards the bound point $\vt_{\mathrm{low}}$ will have the lower noise and lower resolution that are preferred for tasks like low-contrast lesion detection, and 2) the images generated towards the bound point $\vt_{\mathrm{high}}$ will have the higher resolution and higher noise that are preferred for tasks like high-contrast structure inspection. It should be noted that the step size for image generation is determined by the learning rate. Therefore, in theory, one can continuously control the noise-resolution tradeoff by choosing a very small learning rate.

    \vspace{-0.5cm}

\subsection{Materials}
\subsubsection{Datasets}
\paragraph{Training and validation datasets}

We used the publicly accessible X-ray Low-dose CT challenge dataset \cite{RN37}(\mbox{https://www.aapm.org/GrandChallenge/LowDoseCT}) to train the DL-based denoiser. Specifically, the officially released training dataset includes ten patients who were scanned with Siemens CT scanners. To simulate the LDCT image, Poisson noise was first inserted into the acquired helical projection data to reach a noise level that corresponds to $25\%$ of the normal dose. Then, both the original normal-dose projection data and the simulated low-dose projection data were reconstructed into CT images, termed normal-dose CT (NDCT) and LDCT, respectively. Each patient case contains around 600 2D slices. We split these ten patient cases into eight for training and two for validation, so we had 4800 2D slices in the training dataset and 1136 2D slices in the validation dataset. The image size was $512\times512$.

\paragraph{Testing datasets}

To further evaluate the proposed method with external datasets, we employed two different realistic datasets. 

The first dataset was collected based on a cadaver study by using a GE CT scanner at Massachusetts General Hospital. In this study, a cadaver was scanned with four different exposure levels, so the resulting CT images correspond to four different noise indexes (NIs): 10, 20, 30 and 40. All the CT images in this dataset were reconstructed with commercial software under default parameter settings.

The second dataset were based on a realistic patient case of prostate cancer treated with image-guided radiation therapy. This dataset was collected with a cone-beam CT (CBCT) imager integrated with a TrueBeam medical linear accelerator (Varian Medical System, Palo Alto, CA). In detail, this dataset contains six CBCT scans collected in three consecutive treatment fractions. At each fraction, a low-dose CBCT scan was taken followed by a normal-dose CBCT scan. All three normal-dose CBCT scans have an exposure setting of $80\mathrm{mA}\times13\mathrm{mA}$, while the three low-dose CBCT scans have exposure settings of $40\mathrm{mA}\times13\mathrm{mA}$, $20\mathrm{mA}\times13\mathrm{mA}$, and $10\mathrm{mA}\times13\mathrm{mA}$. All six projection datasets were FDK-reconstructed into CBCT images of size $512\times512\times256$ with a voxel size of $1.0\times1.0\times1.0\ \mathrm{mm}^3$. We extracted a representative image for algorithm validation. 

\subsubsection{Experimental Design}
\paragraph{Denoiser in the training phase}

Without loss of generality, in this work, we used the well known U-Net \cite{RN38} to train the denoiser for the purpose of demonstrating the algorithm. Because the images in the training dataset have a size of $512\times512$, we set the downsampling depth at 9, so the bottleneck layer has a feature dimension of $1\times1$. We adopted the feature doubling strategy in the upsampling stages. The initial feature channel was set as 32. With regard to the network modules, we used the instance normalization \cite{RN39} technique as the normalization layer and the rectified linear unit (ReLU) as the nonlinear activation layer. The convolutional operator has a kernel size of $3\times3$. We used a stride two convolution operator to conduct the downsampling operation.

The Adam \cite{RN40} optimizer was employed to minimize the cost function defined in problem~(\ref{eq:supervised_training}) by $1\times{10}^5$ iterations. The initial learning rate was set as $1\times{10}^{-4}$ and was reduced by 10 times at $5\times{10}^4$ and $7.5\times{10}^4$ iterations, so the corresponding learning rates were $1\times{10}^{-5}$ and $1\times{10}^{-6}$, respectively. The two hyperparameters associated with the Adam optimizer were set as $\beta_1=0.9$ and $\beta_2=0.999$, respectively. The batch size was 1.

For clarity, in what follows, we will refer to the denoiser above as the U-Net denoiser. 

\paragraph{Interactive denoising in the testing phase}

As stated previously, given the above pre-trained denoiser, optimizing cost functions~(\ref{eq:towards_low_noise}) and~(\ref{eq:towards_high_resolution}) with different guidance images in the testing phase can produce multiple images that exhibit different noise-resolution tradeoffs. 

For all the experiments in the testing phase, we used the stochastic gradient descent (SGD) with 0.9 momentum. The learning rate was fixed at $1\times{10}^{-2}$. The optimization process was terminated when the relative change between two consecutive outputs was less than $1\%$. 

\paragraph{Experiments}

We first conducted experiments based on the validation dataset, which has the same distribution as the training dataset. In detail, to show the task-specific denoising, we selected two 2D CT images from the validation dataset, one of which contains many high-resolution structures and the other of which contains a low-contrast lesion. We produced the associated default denoised images by feeding these two noisy images into the pre-trained denoiser. Both of the bound images $\vt_{\mathrm{high}}$ and $\vt_{\mathrm{low}}$ were generated. To show the gradually transiting property in terms of the noise-resolution tradeoff during the optimization processes regarding problems~(\ref{eq:towards_low_noise}) and~(\ref{eq:towards_high_resolution}), we selected one image candidate representing the high-resolution direction and one candidate representing the low-noise direction. For comparison, we also show the default denoised image, the original input LDCT and the reference NDCT. 

We calculated the root mean squared error (RMSE) against the associated NDCT for all the images. Moreover, for the low-contrast lesion detection task, we calculated the contrast-to-noise ratio (CNR) of the lesion region of interest (ROI) for quantitative comparison. The CNR is calculated as $\mathrm{CNR}=\frac{2\left|S-S_b\right|}{\sigma+\sigma_b}$, where $S$ and $S_b$ represent the mean intensities of the ROI and the background, respectively, and $\sigma$ and $\sigma_b$ are the associated standard deviations.

We then demonstrated the performance of the proposed method on a realistic dataset based on the cadaver study. For this site, we have four CT slices associated with four different NIs: 10, 20, 30 and 40. Note that these four slices are not perfectly pixelwise matched, since they were scanned at different time points. The proposed method was applied on all four slices. We show two representative image candidates corresponding to the high-resolution and the low-noise directions, as well as the default denoised image and the original input image. We performed the same experiments as on the above-mentioned CBCT datasets that have three different exposure settings.
 
For each of the experiments, we manually selected a flat ROI to calculate the standard deviation (STD).

\subsubsection{Ablation studies}
\paragraph{Different CNN architectures}

In this paper, the default denoiser is based on the widely used U-Net architecture. To verify the generalizability of our DID algorithm across network architectures, we used a simpler but more general architecture to train the denoiser. More specifically, this architecture contains eight consecutive convolutional layers with no downsampling or upsampling operators. Each layer consists of three operators: a convolution operator with a kernel size of $3\times3$, an instance normalization operator and a ReLU operator. As such, the input image size is the same as the output image size. We refer to this network as plain network.

All the other training details were the same as with the U-Net-based denoiser. We conducted similar evaluation experiments as above. In detail, without loss of generality, we used the image associated with high-resolution structures from the validation dataset and the image associated with an NI of 30 from the cadaver study to verify the algorithm. All other experimental settings in the testing phase were the same as above.

\paragraph{Different training datasets with different noise levels}

Finally, to test the generalizability of the proposed DID algorithm across different training datasets with different noise levels, we trained four different denoisers with four different datasets. The differences among these training datasets were the low-dose levels. In detail, in the original training datasets, each training image pair consists of a simulated quarter-dose-level LDCT and the associated NDCT. As such, we can get the associated noise component by subtracting the NDCT from the LDCT. Consequently, we can simulate the half-dose-level, $\frac{1}{8}$-dose-level and $\frac{1}{16}$-dose-level LDCTs by adding back half, twice, and four times the noise into the NDCT, respectively. We then trained four different denoisers with these four (the above three plus the original dataset, where the LDCTs have a quarter-dose level) different datasets corresponding to four different noise levels. The architecture and the training details were the same as with the default U-Net denoiser. We used the image associated with high-resolution structures from the validation dataset as the test image, which has a default $25\%$-dose level compared to the NDCT.
    \vspace{-0.3cm}

\section{Results}

\begin{figure*}
    \centering
    \includegraphics[width=0.8\textwidth]{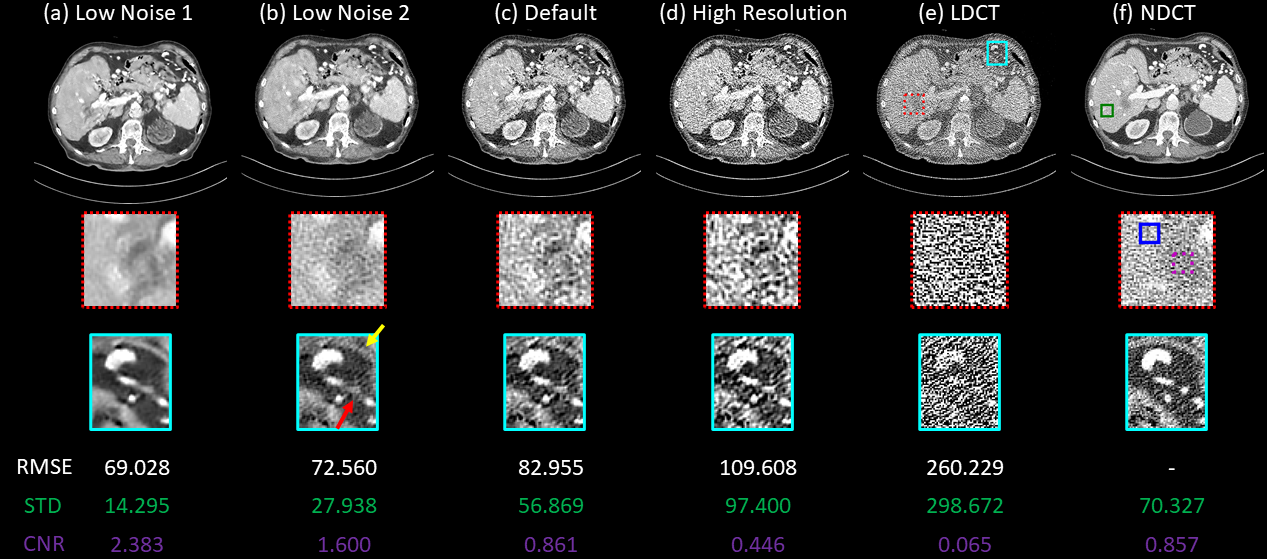}
        \vspace{-0.3cm}

    \caption{Denoised image candidates generated with respect to an image in the validation dataset for low-contrast lesion detection task. From left to right: (a) low noise 1, (b) low noise 2, (c) default, and (d) high resolution image candidates, as well as (e) LDCT and (f) NDCT images. Row 2 shows the zoomed-in views of a low-contrast region of interest (ROI) marked by the red dot box in (e) of Row 1, and Row 3 shows the zoomed-in views of a high-contrast ROI marked by the cyan solid box in (e) of Row 1. The STDs are calculated based on a flat region marked by the green solid box in (f) of Row 1. The contrast-to-noise ratios (CNRs) were calculated based on the foreground content marked by the purple dot box and the background content marked by the blue solid box in (f) of Row 2. The root mean squared error (RMSE) values were calculated against the NDCT. Display window: [-160, 240] HU.}
    \label{fig:AAPM_Lesion_1}
\end{figure*}

Figure~\ref{fig:AAPM_Lesion_1} illustrates the detectability of low-contrast lesions on different denoised image candidates with different noise-resolution tradeoffs generated from an image from the validation dataset. Since the detectability of the low-contrast structure is generally more sensitive to the noise strength, we show two different image candidates (a) and (b) along the low-noise processing direction. We can see by viewing the image candidates from (d) to (a) that the detectability of the low-contrast lesion marked by the red dot box increases as more noise is suppressed. This phenonmenon can be verified in the associated zoomed-in view shown in Row 2 of Figure~\ref{fig:AAPM_Lesion_1}, where this lesion is hardly perceptible in the noisy LDCT. It should be noted that the default denoised image (c) is still too noisy for this low-contrast detection clinical task. The low-noise image candidate (b) is the most similar to the NDCT image (f) in terms of the lesion’s appearance, while the lesion's detectability is highest in low-noise image candidate (a), which can be more clearly observed from the zoomed-in view. To quantitatively evaluate the detectability of this low-contrast lesion, we calculated the CNR, where the foreground and background contents are specified by the purple dot and the blue solid boxes, respectively, as indicated in the zoomed-in view in Figure~\ref{fig:AAPM_Lesion_1}. As expected, the more noise was suppressed, the higher the resulting CNRs, which suggests gradually increasing detectability. 

It is surprising that the CNR associated with the NDCT image was 0.857, which was comparable to that of the default denoised image, even though the visual quality of the default denoised image is much worse than that of NDCT image and also the low-noise image candidates. One possible reason for this is that their noise strengths might be comparable, but their noise textures are distinctly different. We quantitatively checked the STDs based on a flat region as marked by the green solid box in Figure~\ref{fig:AAPM_Lesion_1}(f), Row 1. The resulting values are listed below each corresponding image. As expected, the STD decreases from high-resolution image candidate (d) to low-noise image candidate (a), which suggests that more and more noise is suppressed. By contrast, it is not surprising that the resolution increases from the low-noise direction to the high-noise direction. 

It should be emphasized that these different noise-resolution tradeoffs do not mean that one image is better than another for all tasks. For example, to inspect the curved structure indicated by the yellow arrow, it might be better to use the low-noise image candidates (a) and (b), because this structure is distorted by the strong streak artifacts stemming from the remaining structured noise in the default denoised image and the high-resolution image candidate. However, to visualize the structure indicated by the red arrow, the default denoised image (c) and the high-resolution image candidate (d) would be perferable, because this structure is blurred in the low-noise direction from the process of overly suppressing the noise. For quantitative comparison, we also calculated the RMSE for all the images by comparing them to the NDCT image. It was interesting to find that the default denoised image does not have the lowest RMSE value. Instead, the image with the lowest noise, image candidate (a), has the smallest RMSE value.

\begin{figure*}
    \centering
    \includegraphics[width=0.8\textwidth]{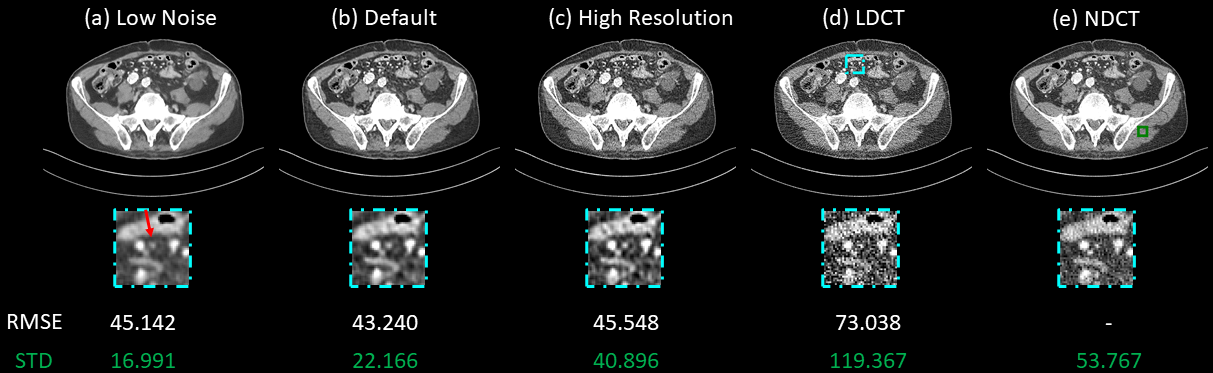}
        \vspace{-0.3cm}

    \caption{Denoised image candidates generated with respect to an image in the validation dataset for high-contrast structure inspection. From left to right: (a) low noise, (b) default, and (c) high resolution image candidates, as well as (d) LDCT and (e) NDCT images. The zoomed-in views below correspond to the region of interest (ROI) marked by the cyan dot dash box in (d) of Row 1. The standard deviations (STDs) were calculated based on a flat region specified by the green solid box in (e) of Row 1. The root mean squared error (RMSE) values were calculated against the NDCT. Display window: [-160, 240] HU.}
    \label{fig:AAPM_HighContrast}
        \vspace{-0.6cm}

\end{figure*}

Figure~\ref{fig:AAPM_HighContrast} presents the image candidates with different noise-resolution tradeoffs generated from the other image from the validation dataset. We assume here that our task is to visualize the high-contrast details as shown in the zoomed-in ROIs. The noise clearly weakens from image (d) to image (a). By contrast, the resolution improves from (a) to (d). As such, one can choose any image candidate with the desired noise-resolution tradeoff to suit the specific clinical task. For example, with regard to the point-like structure indicated by the red arrow, the high-resolution candidate image (c) shows the best discriminativeness, while the default denoised image (b) and the low-noise image candidate (a) both exhibit blurring effects, and the original LDCT (d) is overwhelmed by the strong noise. Furthermore, this gradual transition property can be quantitatively observed from the STDs associated with the selected flat region indicated by the green solid box in Figure~\ref{fig:AAPM_HighContrast}(e). We also calculated the associated RMSE for each processed image against the NDCT image. Despite their differences in resolution, all three denoised images had comparable RMSE values. 

\begin{figure*}
    \centering
    \includegraphics[width=0.8\textwidth]{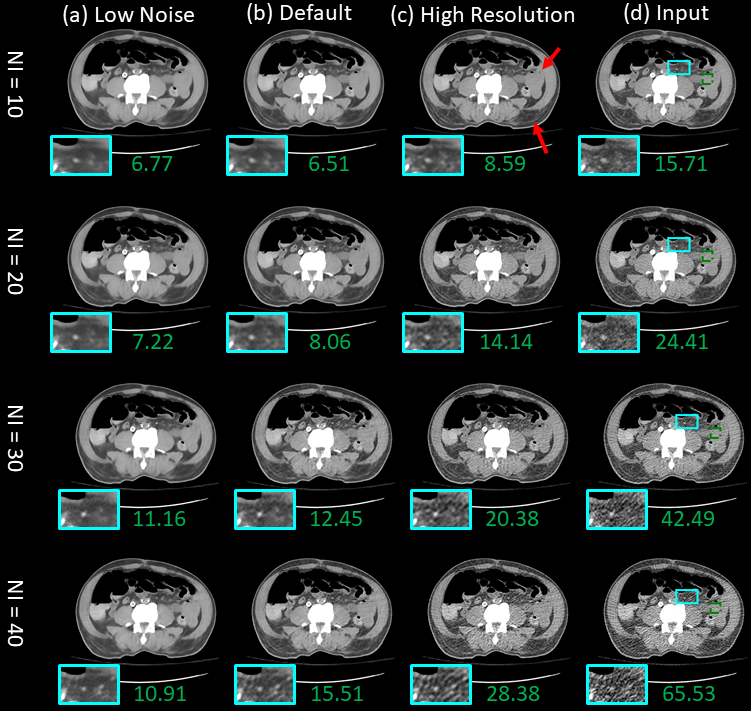}
        \vspace{-0.3cm}

    \caption{Denoised image candidates generated from the cadaver study. From left to right: (a) low noise, (b) default, and (c) high resolution image candidates and (d) input image. From top to bottom, the rows correspond to different noise indexes from 10 to 40. The zoomed-in views correspond to the region of interest (ROI) marked by the cyan solid box. The standard deviations (STDs) were calculated based on a flat region specified by the green dot dash box. Display window: [-160, 240] HU.}
    \label{fig:cadaver}
        \vspace{-0.8cm}

\end{figure*}

Figure~\ref{fig:cadaver} illustrates the denoised results for the images with different NIs from the cadaver study. We can observe that the default model, which was trained to predict the associated NDCT image from the LDCT image of quarter-dose level, produces over-smoothed images when the noise of the input image is already very weak, such as the images corresponding to NIs of 10 and 20. For images that have stronger noise, such as those with NIs of 30 and 40, the processed images exhibit overall good noise-resolution tradeoffs. Because our DID algorithm can produce multiple image candidates, we can restore the overly smoothed structures by generating an image candidate towards the high-resolution direction, as indicated by the red arrow as well as the zoomed-in view in the first row of Figure~\ref{fig:cadaver}. For images that already have overall good noise-resolution tradeoffs, our DID algorithm can still provide other low-noise-preferred or high-resolution-preferred image candidates to adapt to different clinical tasks. For quantitative evaluation, we calculated the STDs for each image candidate based on the specified flat region indicated by the green dash box in each input image. One can observe that, by applying the DID algorithm, the STDs become smaller and smaller as the image candidates generated shift from the high-resolution direction to the low-noise direction.

Similar phenomena can also be observed from the CBCT dataset experiments where multiple images are generated with gradually transiting noise-resolution tradeoffs. See Figure S1 in the supporting document for details.

\begin{figure*}
    \centering
    \includegraphics[width=0.8\textwidth]{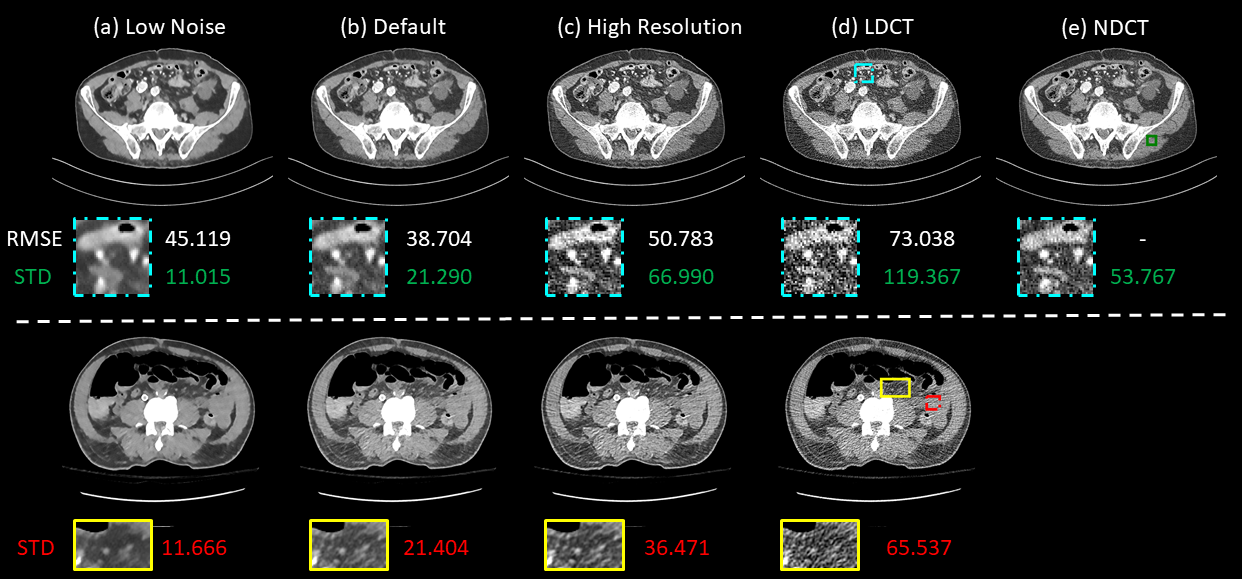}
        \vspace{-0.3cm}

    \caption{Denoised image candidates generated with the plain network. From left to right: (a) low noise, (b) default, and (c) high resolution image candidates, as well as  (d) LDCT and (e) NDCT images. The case above the dashed line corresponds to an image in the validation dataset for high-contrast structure inspection. The case below the dashed line corresponds to an image from cadaver study with a noise index of 40. The root mean squared error (RMSE) values were calculated against the NDCT. The standard deviations (STDs) were calculated based on a flat region specified by the green solid box (above) and the red dot dash box (below). The zoomed-in views correspond to the region of interest (ROI) marked by the cyan dot dash box (above) and the ROI marked by the yellow solid box (below). Display window: [-160, 240] HU.}
    \label{fig:plainnet}
\end{figure*}

Figure~\ref{fig:plainnet} demonstrates our DID algorithm’s performance based on a more general network architecture, i.e., plain network. We can clearly observe the gradual transiting phenomenon with regard to the noise-resolution tradeoff. Specifically, in the case that corresponds to the high-contrast slice from the validation dataset, we can see that the high-resolution image candidate reveals clear structure boundaries, which are either smoothed out in the default image and low-noise image candidate or overwhelmed by the strong quantum noise in the original LDCT image. This phenomenon can be further verified in the zoomed-in views of the ROI marked by the cyan dot dash box. It should be mentioned that the high-resolution image candidate might be more clinically useful for certain tasks that demand high resolution, even though it has the largest RMSE value of all the illustrated image candidates. By visually comparing the high-resolution image candidate with the associated NDCT image, we can see that the former exhibits comparable image resolution but slightly stronger noise. The noise strength comparison can be confirmed quantitatively based on the STDs of the ROI marked by the green solid box. In the case that corresponds to the cadaver study with an NI of 40, all the image candidates have their own values: 1) the low-noise image candidate shows well-suppressed noise but inferior resolution, 2) the high-resolution image candidate shows superior resolution but a stronger streaky artifact, and 3) the default denoised image shows intermediate image properties. This observation can be further supported by the zoomed-in view of the ROI marked by the yellow solid box, as well as the STDs of the ROI marked by the red dot dash box. 

\begin{figure*}
    \centering
    \includegraphics[width=0.8\textwidth]{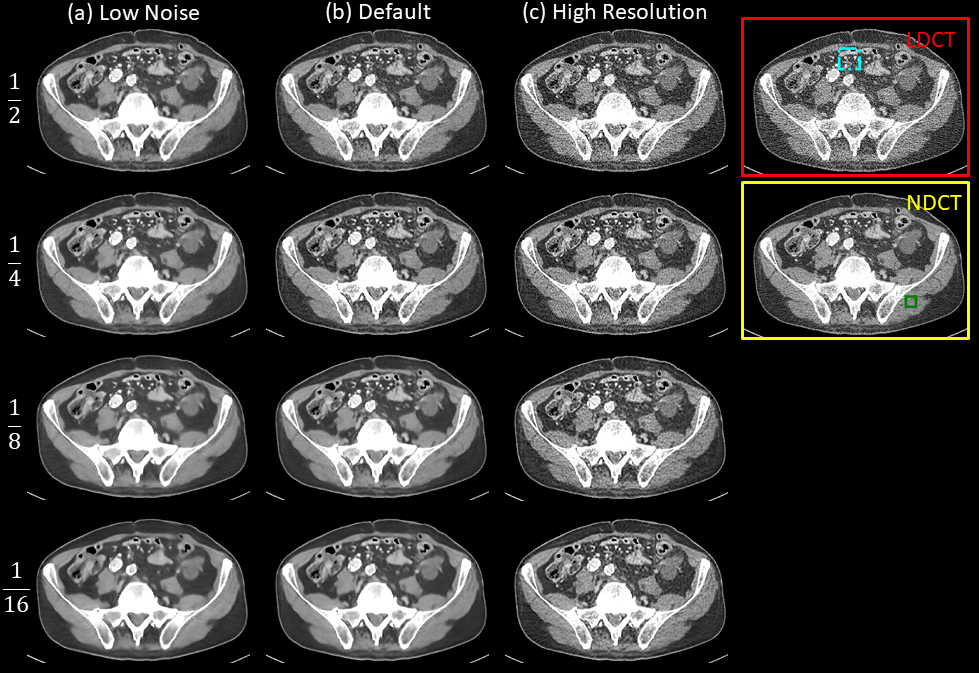}
        \vspace{-0.3cm}

    \caption{Denoised image candidates generated with different training datasets that have different noise levels. From left to right: (a) low noise, (b) default, and (c) high resolution image candidates. The LDCT and NDCT images are presented in the right-most column. From top to bottom, each row corresponds to a different training dataset with a different dose level setting, from $\frac{1}{2}$ to $\frac{1}{16}$ of the NDCT. The testing image has a dose level of $\frac{1}{4}$ of the NDCT. Display window: [-160, 240] HU.}
    \label{fig:train_different_noise_level}
        \vspace{-0.5cm}

\end{figure*}

\begin{figure}
    \centering
    \includegraphics[width=0.5\textwidth]{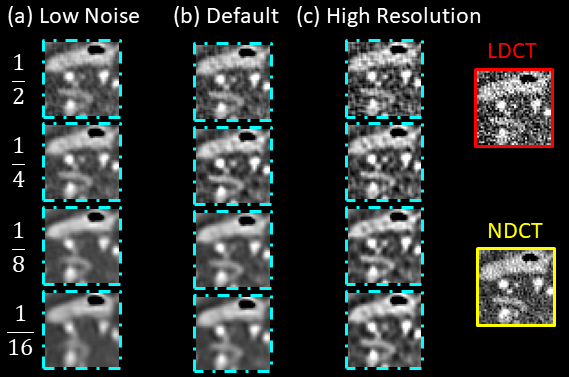}
        \vspace{-0.3cm}

    \caption{Zoomed-in view for the regions of interest (ROIs) marked by the cyan dot dash box in Fig.~\ref{fig:train_different_noise_level}. From left to right: (a) low noise, (b) default, and (c) high resolution image candidates. The LDCT and NDCT images are presented in the right-most column. From top to bottom, rows correspond to different training datasets with dose level settings from $\frac{1}{2}$ to $\frac{1}{16}$ of the NDCT. The testing image has a dose level of $\frac{1}{4}$ of the NDCT. Display window: [-160, 240] HU.}
    \label{fig:train_different_noise_level_roi}
        \vspace{-0.8cm}

\end{figure}

We now show the adaptation of the proposed DID algorithm on the training datasets with different noise levels. Figure~\ref{fig:train_different_noise_level} shows that, as the dose levels of the training datasets decrease from $\frac{1}{2}$ to $\frac{1}{16}$, the default denoised images (b) become smoother and smoother, and the image denoised with the model trained on the $25\%$-dose level training dataset shows the best noise-resolution tradeoffs overall. After we apply the proposed DID algorithm, by selecting a desired image candidate, we can effectively shift this tradeoff to the desired degree for the images processed with models associated with different dose levels. More specifically, for the $\frac{1}{2}$-dose level dataset-based model, users can choose the low-noise image candidate if they feel that the default denoised image is too noisy. Regarding the models trained based on the $\frac{1}{8}$ or $\frac{1}{16}$-dose level datasets, one can choose the high-resolution image candidates to reveal more details. These phenomena can be more clearly observed from the associated zoomed-in view shown in Figure~\ref{fig:train_different_noise_level_roi} for the ROI marked by the cyan dot dash box in  Figure~\ref{fig:train_different_noise_level}. 
    \vspace{-0.5cm}

\section{Discussion and Conclusion}

It is well accepted that the evaluation of medical image quality is highly task-specific, and sometimes even user-specific, depending on users' clinical experience and image appearance preferences. Having benefited from the rapid development of DL techniques, AI-based automation tools have been developed extensively for healthcare. However, there are still obstacles to their clinical deployment. One reason for this might be that AI can usually only provide a general result that reflects the overall properties of the training dataset. Since the final usable result usually must be task-specific and user-specific, further customization of this general result is essential. We believe that human-centered AI in medicine is the way to go for the clinical implementation of AI technologies, where AI will not replace humans by fully automating everything, but rather, will assist humans to perform clinical tasks better and faster. 

With regard to denoising medical images, many excellent DL-based denoisers have been developed. Compared to the conventional methods, these DL-based automatic denoisers show generally superior performance when the training and testing environments are consistent (such as the default denoised images defined in this paper). However, they cannot deliver task-specific and user-specific results, because they usually only produce a single output that reflects the overall properties of the training dataset. Besides, their performance will be decreased when there is a dataset domain shift. Given these challenges, we proposed the DID algorithm to run on top of existing DL-based denoisers to assist users in interactively tuning images to achieve the noise-resolution tradeoff desired for the specific task. 

This is achieved by solving problems~(\ref{eq:towards_low_noise}) and~(\ref{eq:towards_high_resolution}) during the testing phase, as detailed in Section~\ref{sec:methods}. Since our DID algorithm is directly used in the model testing phase, it does not need to interrupt the training pipelines of existing models, so it can be readily combined with any state-of-the-art DL-based denoiser. We have demonstrated this property in this work by evaluating the algorithm’s performance on two different but popular network architectures: U-Net (Figs.~\ref{fig:AAPM_Lesion_1} to~\ref{fig:cadaver}) and plain network (Fig.~\ref{fig:plainnet}).

Moreover, the proposed DID algorithm can also mitigate the problem of model generalizability, which has been widely recognized in DL-based denoisers. Basically, by learning from the training dataset, the DL-based denoisers are trained to suppress the noise of a CT image and are only expected to reach the overall optimal denoising performance (in terms of some predefined metrics, such as RMSE in this paper) when the noise levels of the LDCT images in the training dataset and the testing images are the same. However, in practice, different LDCT scans may have different noise levels due to different anatomical sites, equipment vendors, imaging protocols and so on. It is also likely that different slices from the same CT scan will have different noise levels due to tube current modulation. A model trained to target a specific noise level would not generalize well to handle other scenarios. As demonstrated in Figure~\ref{fig:cadaver}, our U-Net model, which was trained on a simulated training dataset, over-smoothed anatomical structures when the NIs of the input CT images were 10 or 20, but produced denoised images with relatively balanced tradeoffs when the associated NIs were 30 or 40. Similar phenomena can also be observed in the CBCT study shown in Figure S1 in the supporting documents: the model smoothed out the structures when the exposure levels were $\frac{1}{2}$ or $\frac{1}{4}$ but produced balanced results when the exposure level was $\frac{1}{8}$. Generally speaking, if the model is trained to deal with stronger noise, it will tend to overly denoise the image, and thereby render a smoothed-out result when the noise of the input image is weaker than that of the training images, and vice versa. For instance, as shown in Figure~\ref{fig:train_different_noise_level}, the model trained based on $\frac{1}{16}$ ($\frac{1}{2}$)-dose level dataset would result in overly smoothed (slightly noisy) images when the input LDCT has a quarter-dose level. Since our DID algorithm can generate multiple image candidates with different noise-resolution tradeoffs, we expect that it will mitigate this generalizability issue by allowing users to choose an appropriate image candidate with the desired properties instead of the default over/under-denoised image. For example, as shown in Figure~\ref{fig:cadaver}, one could choose the high-resolution image candidates for the input images with NIs of 10 and 20 or the low-noise image candidates for the input image with an NI of 40. 

To enable a smooth and friendly interactive user experience between the model and the physician, computational efficiency plays a vital role. In this work, we use the U-Net as the default architecture with an input size of $512\times512$. It only takes tens of milliseconds per iteration to solve problems~(\ref{eq:towards_low_noise}) and~(\ref{eq:towards_high_resolution}) by using a single NVIDIA Titan X GPU card. This can be considered real-time image generation. Using heavier or lighter model architectures could potentially improve model performance or efficiency. One could also pre-generate all the image candidates and store them on the local drive, so that physicians could browse them without any time delay. 

Aside from the computation efficiency, the user experience could also be improved by quantifying the noise-resolution tradeoff of the resultant image with some simple and understandable metrics. This could greatly help physicians to quickly locate the desired image candidate. This is beyond the scope of this paper, but we will keep it as a potential direction for future research.

Lastly, in this work, we used DL-based denoisers to prove the concept of deep interactive denoising. Theoretically speaking, one could easily extend the proposed DID algorithm to refine the default results for other conventional denoisers.

In summary, our target in this paper is to develop a human-centered AI-based denoiser that can interact with physicians for task-specific denoising. To achieve this goal, we introduced an optimization process into the testing phase that can generate multiple images with various noise-resolution tradeoffs from a default denoised image produced by a pre-trained denoiser. Experimental results involving multiple simulation and realistic patient datasets demonstrated that the proposed algorithm can deliver multiple image candidates with different noise-resolution tradeoffs, which can be used for task-specific applications. The results also showed the superior generalizability of our algorithm, as it can adapt to different architectures and different training and testing datasets with different noise levels. 

\section*{Acknowledgment}
We would like to thank Dr. Chenyang Shen for instructive discussions and Dr. Jonathan Feinberg for editing the manuscript.
\vspace{-0.8em}
\bibliographystyle{IEEEtran}
\bibliography{./DID}

\end{document}